\documentstyle{l-aa}

\newcommand{\aaa}[2]{A\&A #1, #2}

\newcommand{\aj}[2]{AJ #1, #2}
\newcommand{\apj}[2]{ApJ #1, #2}
\newcommand{\apjs}[2]{ApJS #1, #2}

\newcommand{\jcp}[2]{J. Comput. Phys. #1, #2}
\newcommand{\mnras}[2]{MNRAS #1, #2}

\begin{document}

\thesaurus{04 (02.07.1;
               03.13.1;
               03.13.4;
               11.05.2;
               11.11.1;
               11.19.2)}

\title{Dynamical effects of softening in $N$-body simulations \mbox{}\\
       of disc galaxies}

\subtitle{Method and first applications}

\author{Alessandro B. Romeo}

\institute{Onsala Space Observatory,
           Chalmers University of Technology,
           S-43992 Onsala, Sweden\\
           E-mail: romeo@oso.chalmers.se}

\date{Received 18 November 1996 / Accepted 16 December 1996}

\maketitle

\begin{abstract}

Two questions that naturally arise in $N$-body simulations of stellar systems
are:
\begin{enumerate}
\item How can we compare experiments that employ different types of softened
      gravity?
\item Given a particular type of softened gravity, which choices of the
      softening length optimize the faithfulness of the experiments to the
      Newtonian dynamics?
\end{enumerate}
We devise a method for exploring the dynamical effects of softening, which
provides detailed answers in the case of 2-D simulations of disc galaxies and
also solves important aspects of the 3-D problem. In the present paper we
focus on two applications that reveal the dynamical differences between the
most representative types of softened gravity, including certain anisotropic
alternatives. Our method is potentially important not only for testing but
also for developing new ideas about softening. Indeed, it opens a {\it
direct\/} route to the discovery of optimal types of softened gravity for
given dynamical requirements, and thus to the accomplishment of a physically
consistent modelling.


\end{abstract}

\section{Introduction}

In $N$-body simulations of stellar systems the gravitational interaction is
modified for curing the Newtonian divergence at short distances. Basically,
such modifications are introduced through a soft cut-off: the softening
length $s$. However, the precise form in which they are implemented can vary.
Since gravity plays a fundamental role in these systems and the gravitational
interaction is modified precisely where it becomes singular, the dynamical
effects of softening should be well understood when designing experiments and
interpreting their results. This dynamical problem has recently stimulated
considerable interest (e.g., Hernquist \& Barnes 1990; Hernquist \& Ostriker
1992; Kandrup et al.\ 1992; Pfenniger 1993; Pfenniger \& Friedli 1993;
Gurzadyan \& Pfenniger 1994; Romeo 1994, hereafter Paper I; Byrd 1995; Gerber
1996; Merritt 1996; Weinberg 1996; Sommer-Larsen et al.\ 1997; Theis 1997;
see also Goodman et al.\ 1993; Farouki \& Salpeter 1994). For extensive
overviews see the above-mentioned Pfenniger \& Friedli (1993), Gurzadyan \&
Pfenniger (1994) and Paper I.

   In Paper I we have investigated the stability problem in the case of 2-D
models with Plummer softening, which are commonly employed in simulations of
disc galaxies. The basic message is that the effect of softening becomes
strongly artificial for $s\ga\lambda/2\pi$, $\lambda$ being the typical
radial wavelength, which means half an order of magnitude below the expected
value. The major results are summarized in the form of a criterion of
approximate physical consistency for $s$ and a stability criterion for the
Toomre parameter. (Other important aspects of the stability problem have been
considered by Byrd 1995.)

   In the present paper we carry out five extensions, as is discussed below.
\begin{enumerate}
\item We generalize the stability analysis of Paper I to an {\it arbitrary\/}
      isotropic form of softening. This is a natural extension since types of
      softened gravity different from the standard Plummer softening are
      becoming more and more commonly employed (e.g., Combes et al.\ 1990;
      Palou\v{s} et al.\ 1993; Shlosman \& Noguchi 1993). In particular, the
      alternatives proposed by Hernquist \& Katz (1989) and Pfenniger \&
      Friedli (1993) reflect an interesting idea, viz.\ that softening should
      be as localized as possible since there is no clear reason for
      modifying the gravitational interaction at long distances, and it is
      tempting to explore its dynamical consequences. From a more general
      point of view, this and the following extensions provide the tools for
      comparing experiments that employ different types of softened gravity.
\item We investigate the implications of our stability analysis for the
      classical relaxation problem (Rybicki 1972; White 1988). Relaxation and
      stability%
\footnote{In this and similar contexts, `stability' should be understood in
          the general sense of `stability properties'; it is not implied that
          the system is stable. The same applies to the use of `relaxation'
          and `equilibrium'.}
      are intimately related in self-gravitating systems, and even simple
      treatments reveal their strong coupling through random motions. On the
      other hand, the contribution of velocity dispersion to the relaxation
      time has been understood only in part and, because of that, the
      classical argument favouring the choice of large values of $s$ is
      wrong. We revise this argument and conclude that neither small nor
      large values of $s$ are convenient. Surprisingly, there exists an
      intermediate choice of $s$ that {\it optimizes\/} the `dynamical
      resolution' of the model, i.e.\ its faithfulness in simulating the
      dynamics of 3-D discs with Newtonian gravity, especially in situations
      near to the stability threshold. We identify the optimal
      characteristics, and show how to evaluate them for a given type of
      softened gravity. In addition to investigating this aspect of the
      relaxation problem, we explain how effectively softening reduces noise
      on various scales.
\item We complete the examination of 2-D models with isotropic softening by
      investigating the equilibrium problem for an axisymmetric state with
      epicyclic motions. In particular, we explain how significantly the
      circular speed and related quantities deviate from their Newtonian
      behaviours at various distances from the centre.
\item We consider 3-D models with isotropic softening and examine two
      limiting cases: discs and the simple, yet instructive, Jeans problem.
      An extension to 3D has been encouraged by Friedli (1994) and Junqueira
      \& Combes (1996, see the interesting remarks in Sect.\ 2.2). Real
      stellar systems have several gravitationally interacting components.
      Both $N$-body simulations and theoretical works are forced to use
      simplified models, which do not necessarily provide faithful
      representations of the complexity of such systems. Our motivation is to
      understand the basic differences between the dynamical effects of
      softening in 3D and 2D, in the presence of a single stellar component
      (3-D vs.\ 2-D discs and Jeans problem vs.\ discs). In particular, we
      point up the {\it strong\/} modifications introduced by a homogeneous
      geometry and the absence of rotation.
\item We complete the examination of 3-D models by discussing the basic
      dynamical effects of softening anisotropy. The idea underlying this
      extension is that softening should be anisotropic in simulations of
      stellar systems where significantly higher spatial resolution is
      required along a certain direction, such as in disc galaxies. The
      alternative family of softening recently proposed by Pfenniger \&
      Friedli (1993) reflects such an important idea, and it is tempting to
      explore the dynamical relations between its members. (An analogous idea
      has been discussed in the context of smoothed particle hydrodynamics by
      Shapiro et al.\ 1994, 1996 and Fulbright et al.\ 1995; Hernquist,
      private communication, has remarked that in models with anisotropic
      smoothing there may be a significant tendency for angular momentum not
      to be conserved.)
\end{enumerate}
These extensions all together form a method for exploring the dynamical
effects of softening in $N$-body simulations of stellar systems. Our method
is described in Sect.\ 2, and is structured as in the previous discussion.
The two applications mentioned in the same context are shown in Sect.\ 3 (see
also Appendix A). The conclusions of this paper are drawn in Sect.\ 4, where
we present our contribution in a more general perspective and motivate future
applications.

\section{Method}

\subsection{2-D models with isotropic softening}

\subsubsection{Stability}

In 2-D discs with isotropic softened gravity, a given surface-density
perturbation $\Sigma_1(\vec{R},t)$ induces a potential perturbation
\begin{equation}
\Phi_1(\vec{R},t)=-G\int\!\varphi_s(|\vec{R}-\vec{R}'|)\,
                          \Sigma_1(\vec{R}',t)\,{\rm d}^2\!\vec{R}'\, ,
\end{equation}
where $-Gm\varphi_s(R)$ is the point-mass potential, which depends on the
softening length $s$ [$\varphi_s(R)$ is also known as the softening kernel].
In analysing the Poisson equation (1), we adopt the lowest-order WKBJ
approach, as in Paper I, but with a different albeit asymptotically
equivalent spectral representation. In simple terms, the new feature consists
in considering perturbations with a radial dependence $g_1(R)=\tilde
g_1\,{\rm J}_0(kR)$ rather than $g_1(R)=\hat g_1\,{\rm e}^{ikR}$, where $k$
is the radial wavenumber and ${\rm J}_\nu$ denotes the Bessel function of the
first kind and order $\nu$ (see, e.g., Abramowitz \& Stegun 1972). The
Bessel-Hankel representation is more convenient than the Fourier
representation because it allows factorizing the convolution in the Poisson
equation directly, without any assumption on the form of $\varphi_s(R)$:
\begin{equation}
\tilde\Phi_1=-2\pi G\,\,{\cal H}_0[\varphi_s(R)](k)\,\,\tilde\Sigma_1\, ,
\end{equation}
where ${\cal H}_\nu$ denotes the Hankel transform of order $\nu$ (see, e.g.,
Sneddon 1972; Bracewell 1986):
\begin{equation}
{\cal H}_\nu[g(R)](k)=\int_0^\infty\!g(R)\,{\rm J}_\nu(kR)\,R\,{\rm d}R\, .
\end{equation}

\paragraph{The reduction factor.}

Equation (2) admits of a simple interpretation: the potential perturbation
induced by a given surface-density perturbation is weakened by a factor
\begin{equation}
{\cal S}\equiv\frac{{\cal H}_0[\varphi_s(R)](k)}
                   {{\cal H}_0[\varphi_{\rm N}(R)](k)}\, ,
\end{equation}
$-Gm\varphi_{\rm N}(R)$ being the Newtonian point-mass potential.
Correspondingly, the contribution of self-gravity to the dispersion relation
is weakened through a reduction of the active unperturbed surface density by
the same factor. Thus, ${\cal S}$ provides complete information about the
effect of isotropic softening on the stability of 2-D discs, and its
evaluation is the starting-point of our method. There are two complementary
routes to ${\cal S}$. One is to calculate it analytically from:
\begin{equation}
{\cal S}(|k|s)=|k|\,\,{\cal H}_0[\varphi_s(R)](k)
\end{equation}
(use, e.g., the comprehensive table of integrals by Gradshteyn \& Ryzhik
1994). This equation follows directly from Eq.\ (4), being $\varphi_{\rm
N}(R)=1/R$ and ${\cal H}_0[1/R](k)=1/|k|$. The other is to compute it
numerically from:
\begin{equation}
{\cal S}(|k|s)=1-{\cal H}_1\left[\frac{1}{R^2}-f_s(R)\right](|k|)\, ,
\end{equation}
$-Gm^2f_s(R)$ being the point-mass force (use, e.g., the extensive and
well-documented NAG library; for algorithms of fast Hankel transform see,
e.g., Gueron 1994; van Veldhuizen et al.\ 1994). This equation is more
convenient than the original Eq.\ (5) because the slow decay of the
oscillatory integrand as $R\rightarrow\infty$ is speeded up. Such an
improvement is obtained with simple tricks, viz.\ integrating by parts and
singling out the Newtonian behaviour of $f_s(R)$ at long distances.

\paragraph{The effective scale height.}

Once ${\cal S}(|k|s)$ has been evaluated, the second step is to analyse its
behaviour at small $|k|s$ since stability is basically a large-scale
property. In order to understand the general features of this behaviour, we
can equivalently start from Eq.\ (5) or (6) and use techniques of asymptotic
expansion of integrals (see, e.g., Bender \& Orszag 1978). In Eq.\ (5), we
should first single out the Newtonian $1/R\,$-dependence of $\varphi_s(R)$ at
long distances, and then expand ${\rm J}_0(kR)$. In Eq.\ (6), we can directly
expand ${\rm J}_1(|k|R)$. The result is that ${\cal S}\sim 1-|k|h$, where in
each case
\begin{equation}
h=\int_0^\infty\![1-R\varphi_s(R)]\,{\rm d}R
 =\frac{1}{2}\int_0^\infty\![1-R^2f_s(R)]\,{\rm d}R\, .
\end{equation}
This quantity is positive in types of softened gravity of practical interest,
and has an important dynamical meaning. In fact, a comparison with the
reduction factor of 3-D discs with Newtonian gravity (Shu 1968; Vandervoort
1970; Romeo 1992) shows that softening mimics thickness on large scales and
$h$ has the effect of a scale height, as far as density waves are concerned
(for bending waves see Masset \& Tagger 1996). As an alternative to Eq.\ (7),
$h$ can be evaluated from the conversion factor
\begin{equation}
\frac{h}{s}=-{\cal S}'(|k|s=0)\, .
\end{equation}

\paragraph{Characteristics.}

The third and last step is to extract detailed information concerning the
stability properties. This part of the method has been described in Paper I
and can be generalized without special difficulties. So in the following
discussion we briefly introduce the basic concepts and point out the major
results of the stability analysis. We adopt the basic fluid description with
the following scaling and parametrization:
\begin{equation}
\bar\lambda\equiv\frac{\lambda}{\lambda_{\rm T}}\, ,
\;\;\;\,\mbox{where}\;\;\;\,
\lambda_{\rm T}=\frac{2\pi}{k_{\rm T}}
               \equiv\frac{4\pi^2G\Sigma}{\kappa^2}\, ,
\end{equation}
\begin{equation}
Q\equiv\frac{c\kappa}{\pi G\Sigma}
\;\;\;\,\mbox{(Safronov-Toomre parameter)}\, ,
\end{equation}
\begin{equation}
\eta\equiv sk_{\rm T}
\;\;\;\,\mbox{(softening parameter)}\, ,
\end{equation}
\begin{equation}
\zeta\equiv hk_{\rm T}
\;\;\;\,\mbox{(effective thickness parameter)}\, ,
\end{equation}
where $\lambda$ is the radial wavelength of the perturbation, $\lambda_{\rm
T}$ is the Toomre wavelength, $\Sigma$ is the unperturbed surface density,
$\kappa$ is the epicyclic frequency and $c$ is the planar sound speed. The
stability properties are described by the marginal stability curve, i.e.\ the
dispersion relation for marginally stable perturbations viewed as a curve in
the $(\bar\lambda,Q^2)$ plane for a given $\eta$:
\begin{equation}
Q^2+4\bar\lambda\left[\bar\lambda-{\cal S}(\eta/\bar\lambda)\right]=0\, .
\end{equation}
In particular, the stability level is measured by the effective parameter
\begin{equation}
Q_{\rm eff}=\frac{Q}{\bar Q(\eta)}\, ,
\end{equation}
where the threshold $\bar Q(\eta)$ corresponds to the square root of the
global maximum of the marginal curve, and the typical radial wavelength
corresponds to the location $\bar\lambda_{\rm max}(\eta)$ of this maximum.
The values of $\eta$ and the related quantities that characterize the
stability properties are presented below.
\begin{itemize}
\item The conversion factor $\zeta/\eta=-{\cal S}'(\eta/\bar\lambda=0)$, now
      expressed in dimensionless form, has already been discussed (cf.\ the
      effective scale height).
\item The safety threshold for approximate physical consistency is $\eta_{\rm
      safe}=\frac{1}{5}\,(\zeta/\eta)^{-1}$. For $\eta\la\eta_{\rm safe}$,
      i.e.\ $\zeta\la\frac{1}{5}$, softening mimics the effect of thickness.
      In particular, $\bar Q\approx 1-2\zeta$ and $\bar\lambda_{\rm
      max}\simeq\frac{1}{2}$. For $\eta\ga\eta_{\rm safe}$, softening causes
      artificial stabilization and a moderate degree of `blueshift'.
\item The critical value $\eta_{\rm crit}$ is such that $\bar Q(\eta_{\rm
      crit})=0$. For $\eta\geq\eta_{\rm crit}$, the fundamental meaning of
      velocity dispersion becomes {\it obscure\/} because the stability level
      is no longer actively controlled by $Q_{\rm eff}$. This fact has
      serious dynamical implications: it artificially precludes the
      possibility of simulating regimes of normal spiral structure, which
      require fine-tuned choices of the stability level.
\item The critical radial wavelength is $\bar\lambda_{\rm
      crit}=\bar\lambda_{\rm max}(\eta_{\rm crit})$. Its deviation from
      $\bar\lambda_{\rm max}(0)=\frac{1}{2}$ indicates the sensitivity of
      $\bar\lambda_{\rm max}(\eta)$ for $\eta<\eta_{\rm crit}$.
\end{itemize}

\subsubsection{Relaxation vs.\ stability}

Let us investigate the implications of our stability analysis for the
classical relaxation problem (Rybicki 1972; White 1988; see also Hockney \&
Eastwood 1988). We first sketch the basic ideas. The Rybicki-White relaxation
time $t_{\rm RW}$ is proportional to the softening length $s$, the cube of
velocity dispersion $\sigma$ and the number of computer particles $N$. In
turn, $\sigma$ is proportional to the Safronov-Toomre parameter $Q$. So it
seems that large values of $s$ are convenient, since they result in a long
$t_{\rm RW}$ for given $Q$ and $N$. The idea underlying this argument is that
a given $Q$ corresponds to a constant stability level as $s$ varies. But it
is not so. In fact, the stability threshold $\bar Q$ depends on $s$ and the
level is measured by the effective parameter $Q_{\rm eff}=Q/\bar Q$. Thus
large values of $s$ are not convenient at all, because they correspond to a
low $\bar Q$ and result in a short $t_{\rm RW}$ for given $Q_{\rm eff}$ and
$N$. Furthermore, if $t_{\rm RW}$ is short for both small and large values of
$s$, there must exist an intermediate choice of $s$ that maximizes $t_{\rm
RW}$. Does it satisfy the criterion of approximate physical consistency? The
answer is: Yes, it does. The identification of the optimal characteristics of
2-D discs with isotropic softened gravity is the focal point of our method.

\paragraph{The corrected Rybicki-White relaxation time.}

Originally, $t_{\rm RW}$ was derived by adopting a simple two-body treatment,
and assuming that $f_s(R)$ is of the type: $f_s(R)=0$ for $R\leq s$,
$f_s(R)=1/R^2$ for $R>s$. A generalization of $t_{\rm RW}$ to an arbitrary
isotropic $f_s(R)$ is easy to derive and useful for comparing the effects of
different types of softened gravity. The resulting $t_{\rm RW}$ is equal to
the original one multiplied by a correction factor ${\cal C}$, which is one
of the relaxation characteristics. Referring to the introductory discussion,
we express $t_{\rm RW}$ in a form that splits the various contributions:
\begin{equation}
t_{\rm RW}={\cal C}\cdot\tau(\eta)\,Q_{\rm eff}^3\,N
           \left(\frac{2\pi^3G^2\Sigma^3}{\kappa^5M_{\rm d}}\right)\, ,
\end{equation}
\begin{equation}
{\cal C}s=\left\{\int_0^\infty\![bf_s(b)]^2\,{\rm d}b\right\}^{-1}\, ,
\end{equation}
\begin{equation}
\tau(\eta)=\eta\,\bar Q^3(\eta)\, ,
\end{equation}
where $\tau(\eta)$ measures the `relaxation level' as $\eta$ varies, $M_{\rm
d}$ is the disc mass and $b$ is the impact parameter. In addition, the
original unspecified $\sigma$ is interpreted as the radial velocity
dispersion $c$ (different specifications would only modify the
proportionality factor).


\begin{figure}
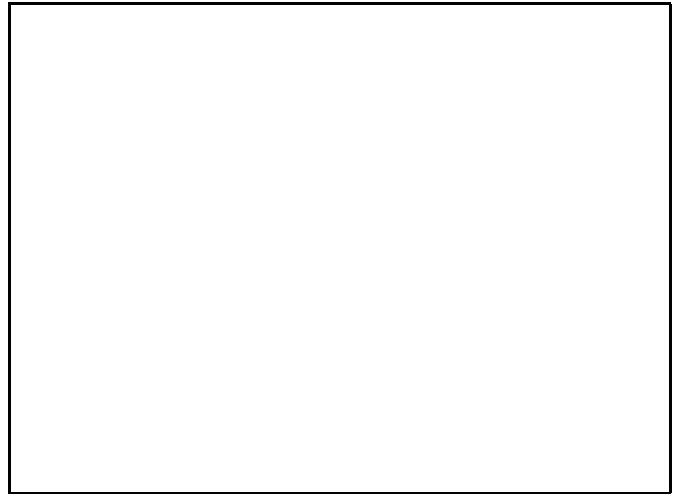

\picplace{6.5cm}
\caption[ ]{The relaxation level $\tau(\eta)=\eta\,\bar Q^3(\eta)$ of 2-D
            discs with Plummer softened gravity, where $\eta$ is the
            softening parameter and $\bar Q(\eta)$ is the stability
            threshold. The contribution of $\bar Q$ to the $\eta$-dependence
            of $\tau$ has major dynamical implications, as is pointed up in
            Sect.\ 2.1.2}
\end{figure}


\paragraph{Optimal characteristics.}

We now identify the values of $\eta$ and the related quantities that optimize
the relaxation and stability properties. Figure 1 shows the behaviour of
$\tau(\eta)$ in the case of Plummer softening, but the following discussion
is general.
\begin{itemize}
\item The optimal relaxation level $\tau_{\rm op}$ and the optimal choice
      $\eta_{\rm op}$ are such that $\tau_{\rm op}=\tau(\eta_{\rm
      op})=\max\,\{\tau(\eta)\}$. The simple analytical approximation
      $\eta_{\rm op}\approx\frac{1}{8}\,(\zeta/\eta)^{-1}$ shows that
      $\eta_{\rm op}<\eta_{\rm safe}$.
\item In 3-D discs with Newtonian gravity, the temperature anisotropy
      corresponding to the stability threshold for the optimal value
      $\zeta_{\rm op}\approx\frac{1}{8}$ would be $\delta_{\rm
      op}\approx\frac{2}{5}$, which means a {\it realistic\/}
      vertical-to-radial velocity dispersion ratio $c_z/c\approx 0.6$ [cf.\
      Paper I, Eq.\ (16) and Fig.\ 3].
\item The values $\eta_{\rm op1}$ and $\eta_{\rm op2}$ such that, e.g.,
      $\tau(\eta_{\rm op1})=\tau(\eta_{\rm op2})=\frac{1}{2}\tau_{\rm op}$
      specify a range of convenient choices of $\eta$, $\eta_{\rm
      op1}\la\eta\la\eta_{\rm op2}$, and complete the information concerning
      the optimal characteristics. The suggested definition is natural and
      especially meaningful because it turns out that $\eta_{\rm
      op2}\sim\eta_{\rm safe}$.
\end{itemize}

\paragraph{What about the reduction factor?}

A delicate aspect of the relaxation problem that has not been considered in
the previous discussion concerns the effects of collective interactions
between particles and self-consistent fluctuations on the dynamical evolution
of the system (e.g., Romeo 1990 and references therein; Weinberg 1993; Zhang
1996). A thorough treatment of collective effects would demand titanic
efforts even in simpler models (cf.\ Weinberg 1993). On the other hand,
useful information is already contained in ${\cal S}(|k|s)$ since the
diffusion properties are determined by the dispersion relation. In
particular, the behaviour of ${\cal S}(|k|s)$ at large $|k|s$ (${\cal S}\ll
1$) shows how effectively softening suppresses small-scale fluctuations,
which represent an important source of noise in 2-D models (cf.\ White 1988;
Schroeder \& Comins 1989).

\subsubsection{Equilibrium}

\paragraph{The reduction factor.}

The equilibrium problem for an axisymmetric state with epicyclic motions can
be solved by using the technique of Hankel transforms, which has already been
introduced in the stability analysis. The basic idea is to ${\cal
H}_0$-transform the Poisson equation twice: once for factorizing the
convolution of $\varphi_s(R)$ and $\Sigma(R)$, and the other time for
recovering $\Phi(R)$ [or, in the inverse problem, $\Sigma(R)$] from its
transform. This is a natural generalization of the approach adopted by Toomre
(1963) in the case with Newtonian gravity, where the Poisson equation can be
expressed in differential form (see also Binney \& Tremaine 1987). It follows
that complete information about the effect of isotropic softening on the
equilibrium of 2-D discs is already contained in ${\cal S}(|k|s)$, which in
this context has the meaning of a reduction factor for the transformed
surface density.

\paragraph{The relative quadratic deviations.}

The formulae for the angular speed $\Omega(R)$ and $\kappa(R)$ are of
particular interest since these quantities also determine the stability and
relaxation properties:
\begin{equation}
\Omega^2(R)=\frac{2\pi G}{R}\,\,
            {\cal H}_1\left[\tilde\Sigma(k)\,{\cal S}(|k|s)\right](R)\, ,
\end{equation}
\begin{eqnarray}
\kappa^2(R) &=& 2\pi G\,\,
                {\cal H}_0\left[k\,\tilde\Sigma(k)\,{\cal S}(|k|s)\right](R)
                \nonumber \\
            & & \mbox{}+\frac{4\pi G}{R}\,\,
                {\cal H}_1\left[\tilde\Sigma(k)\,{\cal S}(|k|s)\right](R)\, ,
\end{eqnarray}
$\tilde\Sigma(k)$ being an abbreviation for ${\cal H}_0[\Sigma(R)](k)$. The
relative quadratic deviations of $\Omega(R)$ and $\kappa(R)$ from their
Newtonian behaviours
\begin{equation}
\epsilon_g(R)=\frac{\left|g^2(R)-g_{\rm N}^2(R)\right|}{g_{\rm N}^2(R)}
\;\;\;\,(g=\Omega,\kappa)
\end{equation}
depend on the model of mass distribution and on $s/R_{\rm d}$, $R_{\rm d}$
being the disc scale length. In order to estimate the magnitude of these
deviations, we can set ${\cal S}\sim 1-|k|h$ and find that $\epsilon_g={\rm
O}(h/R_{\rm d})\ll 1$, which is strictly valid for $R\gg h$ [the approximate
formula for the circular speed is: $v_{\rm c}^2(R)\approx v_{\rm
cN}^2(R)+2\pi GR\Sigma'(R)h$]. More important are the deviations that result
near the centre. They imply a change in the number and/or location of the
inner Lindblad resonances, for a given pattern speed, together with larger
natural scales for $\lambda$, $c$, $s$, $h$ and $t_{\rm RW}$. The
contribution of a massive bulge to the rotation curve makes the system more
robust against such modifications, but in certain respects this modelling is
unclear and we are not yet in a position to draw quantitative conclusions
(cf.\ Bertin 1996; Bertin \& Lin 1996; Junqueira \& Combes 1996, Sect.\ 2.2).


\begin{figure*}
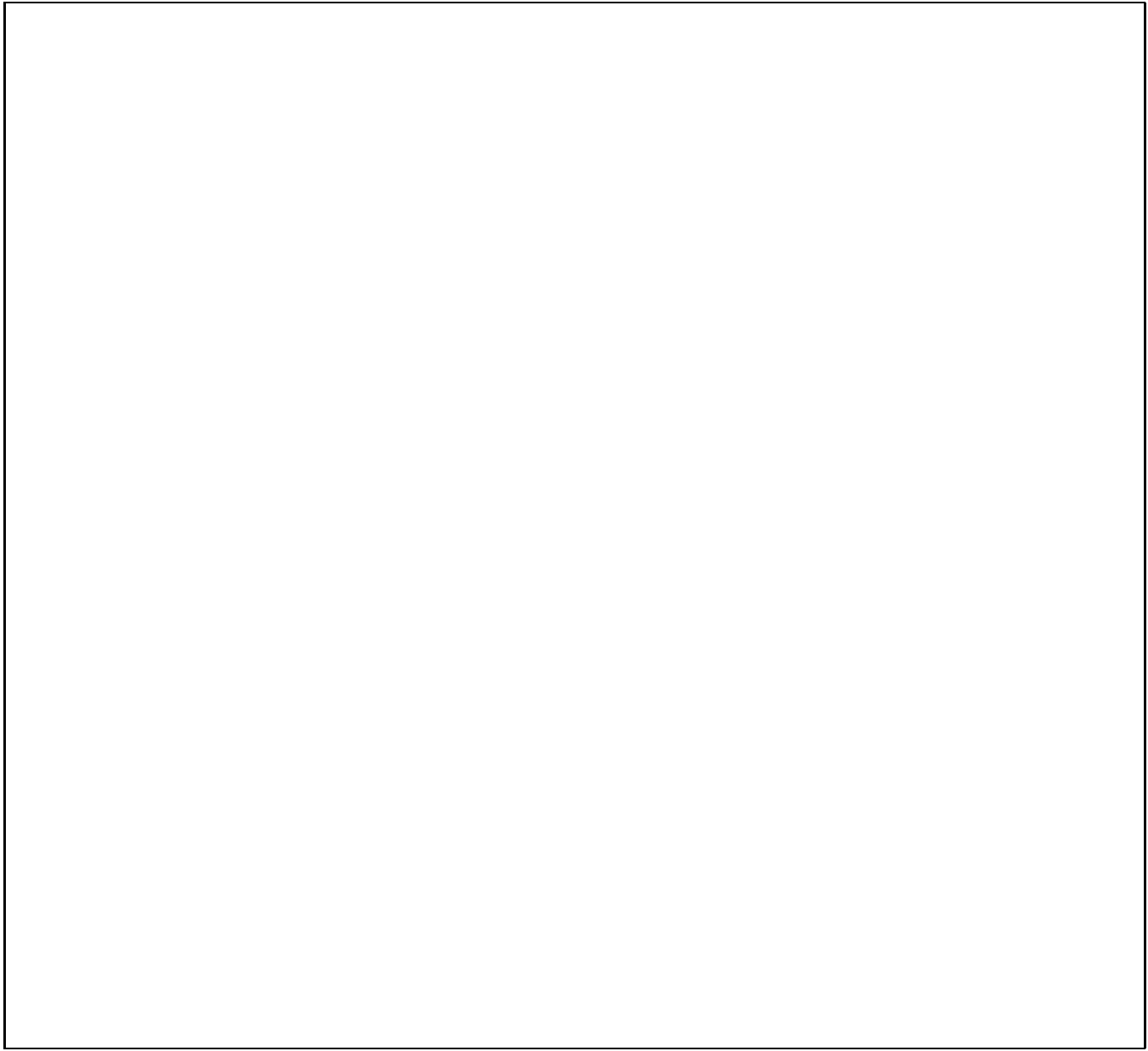

\picplace{16.5cm}
\caption[ ]{The point-mass potential $-Gm\varphi(\vec{r})$ (top) and the
            magnitude of the force $Gm^2|\vec{f}|(\vec{r})$ (bottom) in
            isotropic (left) and anisotropic (right) types of softened
            gravity. In the isotropic types (discussed in Sect.\ 3.1) the
            gravitational interaction depends on the softening length $s$. In
            the anisotropic types (discussed in Sect.\ 3.2) the gravitational
            interaction depends on the planar and vertical softening
            semi-axes $s_\parallel$ and $s_\perp$, respectively. The
            anisotropic behaviour is only shown in the extreme cases
            $(R,|z|=0)$ and $(R=0,|z|)$, and is labelled accordingly}
\end{figure*}


\begin{figure*}
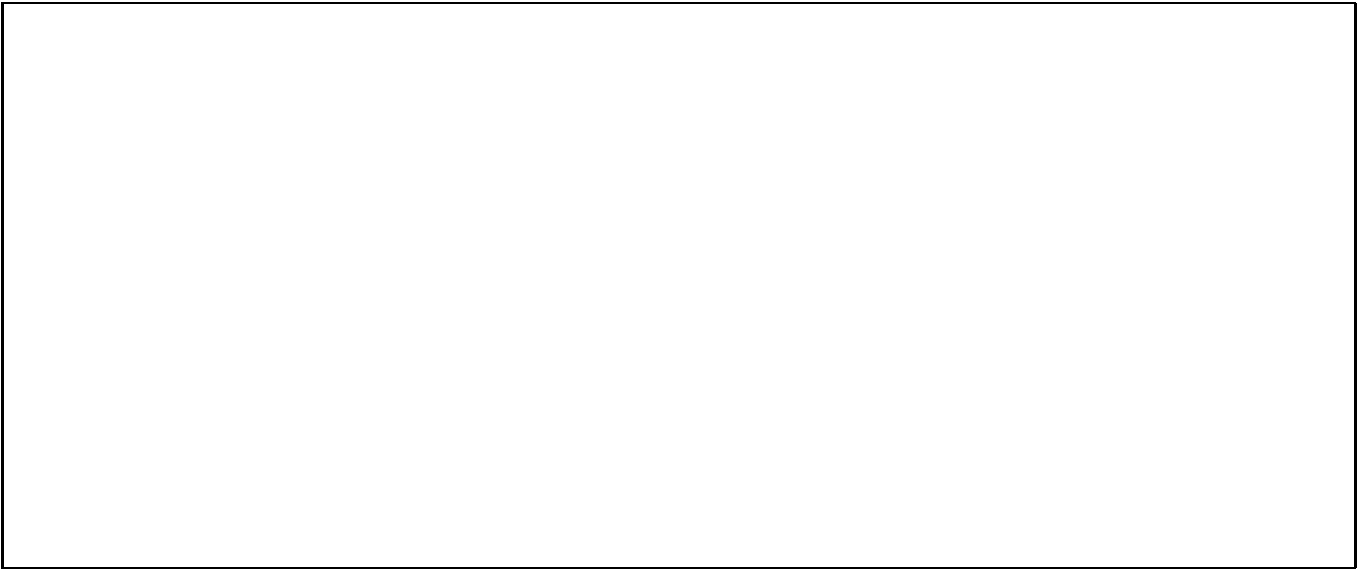

\picplace{7.5cm}
\caption[ ]{The reduction factor ${\cal S}(k)$ of 2-D discs with isotropic
            (left) and anisotropic (right) types of softened gravity, $k$
            being the radial wavenumber of the perturbation. In addition, $s$
            is the softening length, $s_\parallel$ is the planar softening
            semi-axis, and the types of softened gravity are abbreviated as
            in Fig.\ 2. Also shown are the ranges in which the contribution
            of self-gravity to the dispersion relation is stabilizing and
            destabilizing. (In the case with Newtonian gravity ${\cal S}=1$)}
\end{figure*}


\begin{table*}
\caption[ ]{Stability and relaxation characteristics of 2-D discs with
            isotropic (top) and anisotropic (bottom) types of softened
            gravity}
\small
\begin{tabular}{llllllllllllll}
\hline
\noalign{\smallskip}
                                                  &
                                                  &
\                                                 &
\multicolumn{5}{l}{\sc Softening Parameter}       &
\                                                 &
\multicolumn{5}{l}{\sc Related Quantities}        \\
\noalign{\smallskip}
\cline{4-8} \cline{10-14}
\noalign{\smallskip}
{\sc Type of Softened Gravity}            &
{\sc Abbr.}                               &
                                          &
${\eta_{\rm op1}}^{\rm a}$                &
${\eta_{\rm op}}^{\rm b}$                 &
${\eta_{\rm op2}}^{\rm a}$                &
${\eta_{\rm safe}}^{\rm c}$               &
${\eta_{\rm crit}}^{\rm d}$               &
                                          &
${\cal C}^{\rm e}$                        &
${\tau_{\rm op}}^{\rm f}$                 &
${\delta_{\rm op}}^{\rm g}$               &
${\zeta/\eta}^{\rm h}$                    &
$\mbox{$\bar\lambda_{\rm crit}$}^{\rm i}$ \\
\noalign{\smallskip}
\hline
\noalign{\smallskip}
Plummer                                          & P   & &
0.03 & 0.12 &     0.25     &     0.20     & 0.37       & &
1.70 & 0.05 &     0.39     &     1.00     & 0.37         \\
Cubic Spline                                     & CS  & &
0.06 & 0.23 &     0.44     &     0.45     & 0.61       & &
0.99 & 0.11 &     0.30     &     0.44     & 0.46         \\
Homogeneous Sphere                               & HS  & &
0.07 & 0.27 &     0.52     &     0.53     & 0.71       & &
0.83 & 0.12 &     0.30     &     0.38     & 0.47         \\
\noalign{\smallskip}
\hline
\noalign{\smallskip}
Homogeneous Oblate Spheroid $^{\rm j}$           & HOS & &
0.18 & 0.55 &     0.94     & ---$\!\!$--- & 1.21       & &
0.44 & 0.32 & ---$\!\!$--- &     0.04     & 0.61         \\
Homogeneous Prolate Spheroid $^{\rm j}$          & HPS & &
0.03 & 0.11 &     0.24     &     0.18     & 0.37       & &
1.73 & 0.05 &     0.39     &     1.13     & 0.33         \\
\noalign{\smallskip}
\hline
\noalign{\medskip}
\end{tabular}
\begin{list}{}{}
\item[$^{\rm a}$] Values corresponding to one half the optimal relaxation
                  level.
\item[$^{\rm b}$] Optimal choice.
\item[$^{\rm c}$] Safety threshold for approximate physical consistency.
\item[$^{\rm d}$] Critical value.
\item[$^{\rm e}$] Correction factor for the Rybicki-White relaxation time.
\item[$^{\rm f}$] Optimal relaxation level.
\item[$^{\rm g}$] Temperature anisotropy corresponding to the stability
                  threshold for the optimal value of the effective thickness
                  parameter.
\item[$^{\rm h}$] Conversion factor, $\zeta$ being the effective thickness
                  parameter (below the safety threshold for approximate
                  physical consistency).
\item[$^{\rm i}$] Critical radial wavelength.
\item[$^{\rm j}$] HOS has $s_\perp/s_\parallel=\frac{1}{10}$ and HPS has
                  $s_\perp/s_\parallel=3$, where $s_\parallel$ and $s_\perp$
                  are the planar and vertical softening semi-axes,
                  respectively, and $s_\parallel$ is regarded as the
                  softening length of reference.
\end{list}
\end{table*}


\begin{figure*}
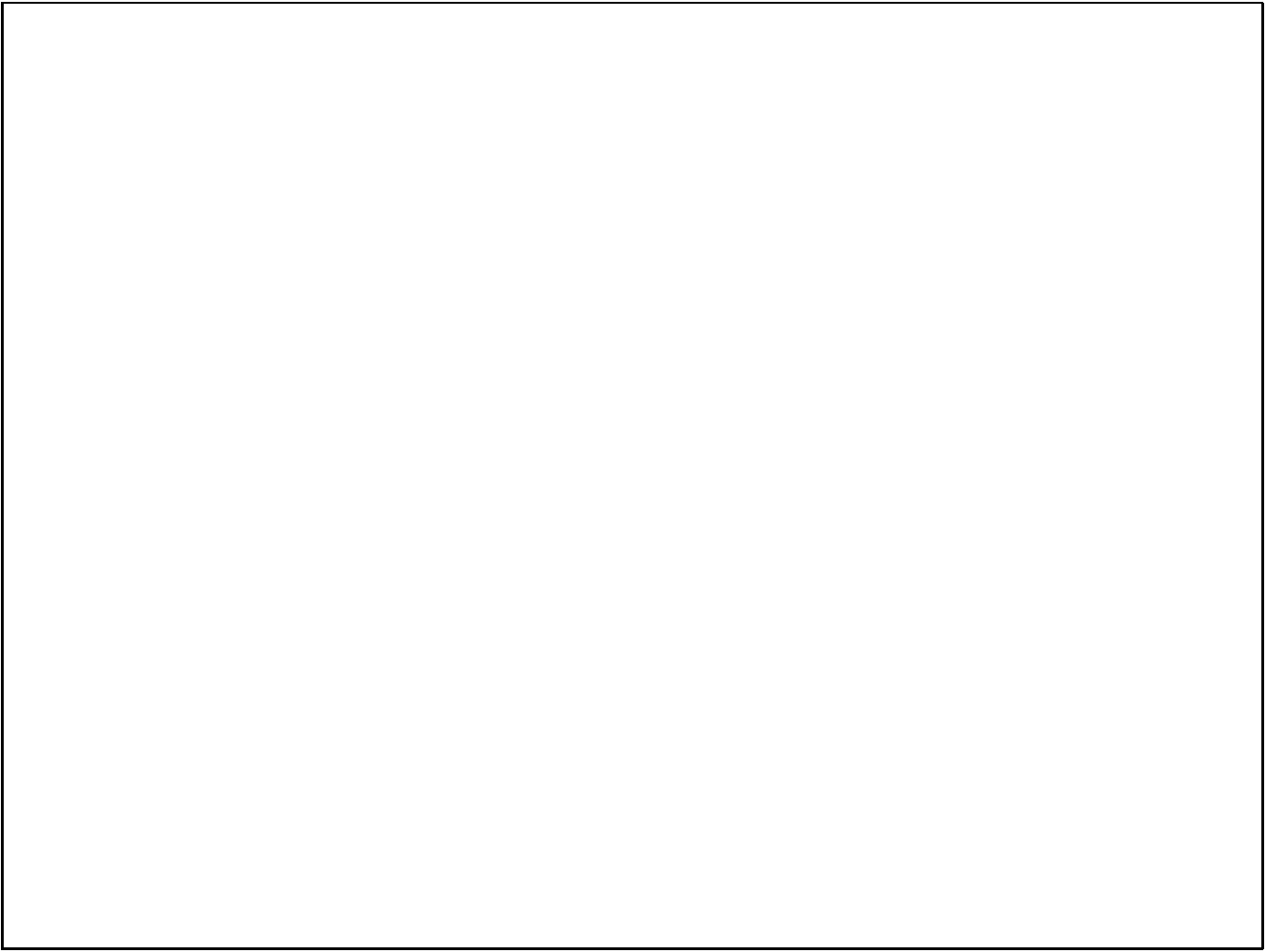

\picplace{13.5cm}
\caption[ ]{The relative quadratic deviations $\epsilon_\Omega(R)$ (top) and
            $\epsilon_\kappa(R)$ (bottom) of 2-D exponential discs with
            isotropic (left) and anisotropic (right) types of softened
            gravity, where $\Omega$ is the angular speed, $\kappa$ is the
            epicyclic frequency and the deviations are measured from their
            Newtonian behaviours. In addition, $R_{\rm d}$ is the disc scale
            length, $s$ is the softening length, $s_\parallel$ is the planar
            softening semi-axis, and the types of softened gravity are
            abbreviated as in Fig.\ 2. (In the case with Newtonian gravity
            $\epsilon_\Omega=\epsilon_\kappa=0$)}
\end{figure*}


\subsection{3-D models with isotropic/anisotropic softening}

\subsubsection{3D vs.\ 2D}

The dynamics of 3-D discs with isotropic softened gravity is difficult to
investigate because the effects of softening combine with those of vertical
random motion in a complicated form. Nevertheless, we do in part understand
how to benefit from the additional degree of freedom introduced into such
models. A simple scale argument suggests that we should choose values of $s$
sufficiently smaller than the Newtonian characteristic scale height,
otherwise softening would significantly affect the vertical structure at
equilibrium, i.e.\ {\it both\/} the mass distribution {\it and\/} the
thickness scale. Likewise, the 2-D stability analysis suggests that in 3D we
should choose $s\ll s_{\rm safe}$, otherwise softening would significantly
interfere with the effect of thickness, as far as density waves are
concerned. Note that choices of $s\geq s_{\rm crit}$ have the same
consequences as in 2D, since the critical stability characteristics do not
depend on the temperature anisotropy. We mention that the vertical structure
at equilibrium and the stability of 3-D discs with Newtonian gravity have
been investigated in previous papers (Romeo 1990, 1992; see also Paper I).

   The 3-D Jeans problem with isotropic softened gravity can be solved by
using the technique of Fourier transforms (see, e.g., Sneddon 1972; Bracewell
1986). In Cartesian coordinates, this is the natural approach for factorizing
the convolution in the Poisson equation and investigating the dynamical
effects of softening, as Pfenniger \& Friedli (1993) have previously
emphasized. In the following discussion we concentrate on the reduction
factor and compare the Jeans problem with the dynamics of discs. The
reduction factor for the volume density can be calculated analytically from:
\begin{equation}
{\cal S}_{\rm J}(|\vec{k}|s)
=|\vec{k}|\,\,{\cal F}_{\rm s}[r\varphi_s(r)](|\vec{k}|)\, ,
\end{equation}
where $\vec{k}$ is the wavevector of the perturbation and ${\cal F}_{\rm s}$
denotes the Fourier sine transform:
\begin{equation}
{\cal F}_{\rm s}[g(r)](|\vec{k}|)
=\int_0^\infty\!g(r)\,\sin(|\vec{k}|r)\,{\rm d}r\, .
\end{equation}
Equation (21) follows from the definition of ${\cal S}_{\rm J}$ and the
spherical symmetry of $\varphi_s$, which allows expressing the 3-D Fourier
exponential transform in terms of ${\cal F}_{\rm s}$. Two alternative
equations that are more convenient for numerical computation can be obtained
by first singling out the Newtonian behaviour of $\varphi_s(r)$ at long
distances, and then integrating by parts (the integrand can be split in two
ways). As regards the magnitude of ${\cal S}_{\rm J}(|\vec{k}|s)$, ${\cal
S}_{\rm J}=1-{\rm o}(|\vec{k}|s)$ at small $|\vec{k}|s$ and ${\cal S}_{\rm
J}\ll 1$ at large $|\vec{k}|s$. The stability properties are weakly affected
by softening. In particular, significant deviations of the marginally stable
wavelength $\lambda_0$ from the Jeans wavelength $\lambda_{\rm J}$ would
occur for values of $s$ comparable to $\lambda_{\rm J}$. For example, in the
case of Plummer softening this can be shown by using the simple numerical
approximation $\lambda_0\approx\lambda_{\rm J}+\frac{3}{2}s$ for
$\frac{1}{6}\lambda_{\rm J}\la s\la\frac{2}{3}\lambda_{\rm J}$ [the
analytical formula for ${\cal S}_{\rm J}(|\vec{k}|s)$ is reported in Appendix
A]. Weinberg (1993) has shown that in analogous models with Newtonian gravity
the dominant source of noise is represented by fluctuations on scales of
stability interest. This fact and our stability analysis suggest that the
collective relaxation properties are also weakly affected by softening. Thus,
there is a {\it sharp\/} contrast between the Jeans problem and the dynamics
of discs: a flattened geometry strengthens the effects of softening, and
rotation makes them critical for values of $s$ that are an order of magnitude
below $\lambda_{\rm T}$.

\subsubsection{Anisotropy vs.\ isotropy}

The dynamical properties of 3-D discs are roughly decoupled parallel and
perpendicular to the plane, and a satisfactory modelling with isotropic
softening may impose significantly different requirements on $s$. Pfenniger
\& Friedli (1993) have discussed an ingenious way of coping with this
difficulty, viz.\ to introduce a further degree of freedom into such models:
softening anisotropy. We assume that the point-mass potential
$-Gm\varphi(\vec{r})$ is of the form $\varphi(\vec{r})=\varphi(R,|z|)$, and
depends on the planar and vertical softening lengths $s_\parallel$ and
$s_\perp$, respectively. The particular form in which softening anisotropy is
implemented defines $s_\parallel$ and $s_\perp$, and specifies their meaning
in the context of the dynamical requirements.

   The 3-D Jeans problem shows the basic dynamical effects of softening
anisotropy. The reduction factor may be evaluated from (cf.\ 2-D discs with
isotropic softened gravity):
\begin{equation}
{\cal S}_{\rm J}(k_\parallel,k_\perp)
=(k_\parallel^2+k_\perp^2)\,\,
 {\cal F}_{\rm c}\left[\tilde\varphi(k_\parallel;|z|)\right](k_\perp)\, ,
\end{equation}
where $\tilde\varphi(k_\parallel;|z|)$ is an abbreviation for ${\cal
H}_0[\varphi(R,|z|)](k_\parallel)$ and ${\cal F}_{\rm c}$ denotes the Fourier
cosine transform. This equation follows from the axial and planar symmetries
of $\varphi$, which allow expressing the 3-D Fourier exponential transform in
terms of ${\cal H}_0$ and ${\cal F}_{\rm c}$. Supposing that
$s_\parallel>s_\perp$, then we expect that ${\cal S}_{\rm
J}(k_\parallel,0)<{\cal S}_{\rm J}(0,k_\perp)$ for $k_\parallel=k_\perp$,
which means that the stability and collective relaxation properties are more
affected parallel than perpendicular to the plane of reference.

\section{Applications}


\begin{figure}
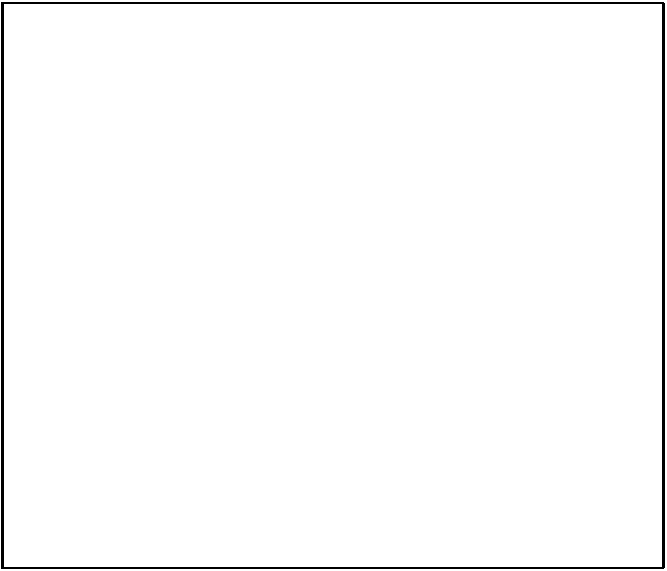

\picplace{7.5cm}
\caption[ ]{The reduction factor ${\cal S}_{\rm J}(|\vec{k}|s)$ for the 3-D
            Jeans problem with isotropic types of softened gravity, where
            $\vec{k}$ is the wavevector of the perturbation and $s$ is the
            softening length. In addition, the types of softened gravity are
            abbreviated as in Fig.\ 2. Also shown are the ranges in which the
            contribution of self-gravity to the dispersion relation is
            stabilizing and destabilizing. (In the case with Newtonian
            gravity ${\cal S}_{\rm J}=1$)}
\end{figure}


\subsection{Isotropic types of softening}

As a first application of our method, we compare the dynamical effects of
three isotropic types of softened gravity: Plummer softening (Aarseth 1963;
Miller 1970), cubic-spline softening (Hernquist \& Katz 1989) and
homo\-geneous-sphere softening (Pfenniger \& Friedli 1993); hereafter
abbreviated to P, CS and HS, respectively. Each name means that the softened
point-mass potential can be viewed as the Newtonian potential generated by a
spherical mass distribution of that type (but the force is evaluated
regarding the test particle as a point mass). In the following discussion we
do not refer to this finite-sized particle interpretation of softening,
unless otherwise specified. The behaviours of these types of softened gravity
are shown in Fig.\ 2 (left). In P the gravitational interaction is softened
mainly at short distances and the Newtonian behaviour is recovered
asymptotically, whereas in CS and HS softening is perfectly localized, i.e.\
beyond a certain cut-off distance the gravitational interaction is
identically Newtonian. The form in which localization is implemented differs
in the two types. In particular, CS has a rather soft cut-off at $r=2s$, $s$
being the nominal softening length, whereas HS has a sharper cut-off at
$r=s$. The results of the dynamical comparison are presented in Fig.\ 3
(left) and Table 1 (top), which are discussed below, and in Figs.\ 4 (left)
and 5. Useful analytical formulae are reported in Appendix A.

   Figure 3 (left) shows that the effective scale height $h$ varies
significantly with softening type, but apart from that the stability
properties are analogous on scales larger than $2h$. This can be shown by
rescaling $k$ in terms of $h^{-1}$ and noting that ${\cal S}$ is similar in
the three types for $|k|h\la\frac{1}{2}$. For $|k|s\ga 4$, ${\cal S}_{\rm
HS}$ changes sign and starts to oscillate, in contrast to ${\cal S}_{\rm P}$.
An analogous behaviour occurs in ${\cal S}_{\rm CS}$, but is less noticeable.
Negative values of ${\cal S}$ mean that a given surface-density perturbation
is in phase with the induced potential perturbation and, correspondingly,
that the contribution of self-gravity to the dispersion relation is
stabilizing. However, this feature does not affect the stability properties
on scales comparable to the inverse of the typical radial wavenumber, since
${\cal S}$ changes sign well beyond the critical point $k_{\rm crit}s_{\rm
crit}$ [cf.\ Table 1 (top)]. The related oscillations of ${\cal S}$ mean that
in certain ranges of $|k|s$ noise is suppressed more effectively on larger
than smaller scales, and correspond to oscillations of the short-wave branch
of the dispersion relation. On the other hand, this feature has not
significant consequences for the relaxation properties, because the corrected
optimal relaxation level ${\cal C}\tau_{\rm op}$ is approximately the same as
in P [cf.\ Table 1 (top)].

   Table 1 (top) shows that the stability and relaxation characteristics vary
significantly with softening type. The largest variations occur between HS
and P, concern $\zeta/\eta$ and $\eta_{\rm safe}$, and are up to a factor of
3. In contrast, the corrected optimal relaxation level ${\cal C}\tau_{\rm
op}$, $\delta_{\rm op}$ and $\bar\lambda_{\rm crit}$ remain approximately
constant. The variations between HS and CS are within 20\%.

\subsection{Anisotropic types of softening}

As a second application of our method, we consider an anisotropic
generalization of HS, the family of homo\-geneous-ellipsoid softening
(Pfenniger \& Friedli 1993), and compare the dynamical effects of two
representative spheroidal members: one oblate with
$s_\parallel:s_\perp=10:1$, $s_\parallel$ and $s_\perp$ being the planar and
vertical softening semi-axes, and the other prolate with
$s_\parallel:s_\perp=1:3$; hereafter abbreviated to HOS and HPS,
respectively. Regarding $s_\parallel$ as the softening length of reference,
$s_\perp/s_\parallel$ gives a measure of softening anisotropy. The behaviours
of these types of softened gravity are shown in Fig.\ 2 (right). Anisotropy
is implemented in a form consistent with the finite-sized particle
interpretation of softening, i.e.\ through a spheroidal deformation of the
field particle. This mainly corresponds to a spheroidal transformation of the
surface on which the force peaks, but it also strongly influences the
behaviour in the plane. In particular, the degree of softening localization,
the sharpness and magnitude of the force peak differ from those in HS [cf.\
Fig.\ 2 (left)], and are higher in HOS than HPS. So a 2-D analysis is
required even though the isotropic case has already been investigated. The
results of the dynamical comparison are presented in Fig.\ 3 (right) and
Table 1 (bottom), which are discussed below, and in Fig.\ 4 (right). The
major points are then generalized in the final discussion. Useful analytical
formulae are reported in Appendix A.

   Figure 3 (right) shows that, for $|k|s_\parallel\la 2$, ${\cal S}_{\rm
HOS}$ is concave, in contrast to ${\cal S}_{\rm HPS}$. The transition
behaviour occurs in ${\cal S}_{\rm HS}$ [cf.\ Fig.\ 3 (left)]. Concavity of
${\cal S}$ mainly means that softening has a stronger tendency to cause
artificial stabilization for a given effective scale height $h$, for we know
that the reduction factor of 3-D discs with Newtonian gravity is convex (cf.\
Paper I, Fig.\ 1). In fact, HOS does not mimic the effect of thickness for
realistically large values of $h$, since the safety threshold for approximate
physical consistency $s_{\rm safe}$ and the optimal value $h_{\rm op}$ are
ill-defined [cf.\ Table 1 (bottom)]. Concavity of ${\cal S}$ also means that
softening causes a moderate degree of `redshift'. For $|k|s_\parallel\ga 4$,
${\cal S}_{\rm HOS}$ and ${\cal S}_{\rm HPS}$ become analogous to ${\cal
S}_{\rm HS}$, and analogous conclusions can be drawn concerning the stability
and relaxation properties.

   Table 1 (bottom) shows that the stability and relaxation characteristics
differ considerably in the two types. The major qualitative differences
concern $\eta_{\rm safe}$ and $\delta_{\rm op}$, and have been pointed out
above [cf.\ discussion of Fig.\ 3 (right)]. Excluding $\zeta/\eta$, which has
a restricted range of applicability in HOS, the largest variation concerns
$\tau_{\rm op}$ and is beyond half an order of magnitude. In contrast, note
the low sensitivity of the corrected optimal relaxation level ${\cal
C}\tau_{\rm op}$, which remains roughly constant.

   Finally, instead of comparing those results with the isotropic case, let
us clarify the dynamical relations between {\it all\/} the spheroidal members
of the family of homo\-geneous-ellipsoid softening and the standard type of
softened gravity [cf.\ Table 1 (top)].
\begin{enumerate}
\item A general feature of the spheroidal members is that $h$ is determined
      by $s_\perp$ alone: $h=\frac{3}{8}s_\perp$. This is not intuitive since
      a more natural softening length in the plane is expected to be
      $s_\parallel$.
\item All the oblate members with $s_\perp/s_\parallel\la\frac{3}{5}$ have an
      ill-defined $s_{\rm safe}$, i.e.\ $s_{\rm safe}\ga s_{\rm crit}$, and
      thus differ fundamentally from P (cf.\ HOS).
\item As $s_\perp/s_\parallel$ increases, such a difference becomes
      progressively less important (cf.\ HS and HPS). Indeed, the prolate
      member with $s_\perp/s_\parallel=\frac{8}{3}$, i.e.\ $h=s_\parallel$,
      turns out to be hardly distinguishable from P. (Values of
      $s_\perp/s_\parallel\ga 2\mbox{--}3$ are not employed in simulations.)
\end{enumerate}

\section{Conclusions}

Modelling gravity is a fundamental problem that must be tackled in $N$-body
simulations of stellar systems, and satisfactory solutions require a deep
understanding of the dynamical effects of softening. This problem has
deserved special attention both in the past (e.g., Miller 1970, 1976) and in
more recent times (e.g., Efstathiou et al.\ 1985; Hernquist \& Katz 1989;
Pfenniger \& Friedli 1993). Viewed in this general perspective, our
contribution has a threefold practical importance in addition to the points
emphasized in Sect.\ 1.
\begin{enumerate}
\item The two present applications of our method reveal the dynamical
      differences between the most representative types of softened gravity:
      the isotropic P, CS and HS (abbreviated as in Sect.\ 3.1), and the
      anisotropic HOS and HPS (abbreviated as in Sect.\ 3.2). The major
      conclusions concerning their dynamical resolution, i.e.\ their
      faithfulness in simulating the Newtonian dynamics, are summarized
      below.
      \begin{enumerate}
      \item As regards the isotropic types, the dynamical resolution is
            comparable. This results from the fact that, even though the
            spatial resolution and the effectiveness in reducing noise differ
            significantly in P, CS and HS for the same nominal softening
            length $s$, those differences can largely be removed by
            considering a more appropriate softening length of reference.
      \item As regards the anisotropic types, the dynamical resolution is
            significantly coupled parallel and perpendicular to the plane. In
            the plane, it decreases in quality from HPS to HOS, and the
            transition occurs in the oblate members for a softening axial
            ratio $s_\perp/s_\parallel\sim\frac{3}{5}$ (HPS is dynamically
            similar to P). These disadvantages result from the finite-sized
            particle implementation of softening anisotropy. On the other
            hand, they have less importance than the advantage of introducing
            such a degree of freedom into 3-D simulations of disc galaxies,
            which has been emphasized by Pfenniger \& Friedli (1993) and in
            our method.
      \item Last but not least, when employing these types of softened
            gravity in simulations of disc galaxies, we should recall that
            the dynamical resolution depends {\it critically\/} on two
            quantities: $s$, or $s_\parallel$ for a given
            $s_\perp/s_\parallel$, and the Safronov-Toomre parameter $Q$
            (cf.\ Paper I). The choice of $s$ or $s_\parallel$ should be
            checked vs.\ the profiles of the characteristic values $s_{\rm
            op}$, $s_{\rm safe}$ and $s_{\rm crit}$, which are tabulated in
            the applications. The choice of $Q$ should be checked vs.\ the
            profiles of the stability threshold $\bar Q$ and level $Q_{\rm
            eff}$, which can be evaluated as is explained in the method.
      \end{enumerate}
\item Our method can be applied for testing new ideas about softening. There
      are two features that encourage such future applications.
      \begin{enumerate}
      \item One is the {\it unified\/} approach adopted for investigating
            stability, relaxation and equilibrium. As a result, full
            information about the dynamical effects of softening is contained
            in a single quantity: the reduction factor ${\cal S}$.
      \item The other is the {\it modular\/} structure of the method. We
            describe step by step how to extract detailed information
            concerning the dynamical properties, starting from ${\cal S}$ and
            pointing out the quantities of major interest.
      \end{enumerate}
\item But our method can be applied in another, more fruitful, way: for
      developing new ideas about softening. Indeed, it opens a {\it direct\/}
      route to the discovery of optimal types of softened gravity for given
      dynamical requirements, and thus to the accomplishment of a physically
      consistent modelling even in the presence of a cold interstellar
      gaseous component. Such a future application will be the objective of a
      `twin' paper.
\end{enumerate}

\begin{acknowledgements}

This paper is dedicated to my parents Francesco and Grazia. It is a great
pleasure to thank the referee, Daniel Pfenniger, whose papers about softening
have represented an invaluable source of inspiration for both developing and
maturing my ideas. In addition, I am very grateful to Francoise Combes and
Lars Hernquist for strong encouragement and valuable suggestions on a
previous draft of this paper. I am also very grateful to John Black, Daniel
Friedli, Cathy Horellou, Henry Kandrup, Richard Miller, Juan Muzzio, Masafumi
Noguchi and Jan Palou\v{s} for strong encouragement and useful discussions.
The financial support of the Swedish Natural Science Research Council is
greatly appreciated.

\end{acknowledgements}

\appendix

\section{Useful analytical formulae for Sect.\ 3}

\begin{equation}
{\cal S}_{\rm P}(|k|s)
={\rm e}^{-|k|s}\, ;
\end{equation}
\begin{equation}
\textstyle
(\zeta/\eta)_{\rm P,\,CS,\,HS,\,HOS,\,HPS}
=1,\,\frac{31}{70},\,\frac{3}{8},\,\frac{3}{80},\,\frac{9}{8}\, ;
\end{equation}
\begin{equation}
\textstyle
(\eta_{\rm safe})_{\rm P,\,CS,\,HS,\,HOS,\,HPS}
=\frac{1}{5},\,\frac{14}{31},\,\frac{8}{15},\,\mbox{---},\,\frac{8}{45}\, ;
\end{equation}
\begin{equation}
\textstyle
(\eta_{\rm crit})_{\rm P}
=\frac{1}{\rm e}\, ;
\end{equation}
\begin{equation}
\textstyle
(\bar\lambda_{\rm crit})_{\rm P}
=\frac{1}{\rm e}\, ;
\end{equation}
\begin{equation}
\textstyle
({\cal C})_{\rm P,\,CS,\,HS}
=\frac{16}{3\pi},\,\frac{17\,325}{17\,504},\,\frac{5}{6}\, ;
\end{equation}
\begin{equation}
{\cal S}_{\rm J\,P}(|\vec{k}|s)
=|\vec{k}|s\,{\rm K}_1(|\vec{k}|s)\, ,
\end{equation}
\begin{equation}
{\cal S}_{\rm J\,CS}(|\vec{k}|s)
=\frac{\sin^3(|\vec{k}|s/2)}{(|\vec{k}|s/2)^3}\,\,
 {\cal S}_{\rm J\,HS}(|\vec{k}|s/2)\, ,
\end{equation}
\begin{equation}
{\cal S}_{\rm J\,HS}(|\vec{k}|s)
=\frac{3\,[\sin(|\vec{k}|s)-|\vec{k}|s\,\cos(|\vec{k}|s)]}
      {(|\vec{k}|s)^3}\, ,
\end{equation}
where ${\rm K}_\nu$ denotes the modified Bessel function of the second kind
and order $\nu$ (see, e.g., Abramowitz \& Stegun 1972).

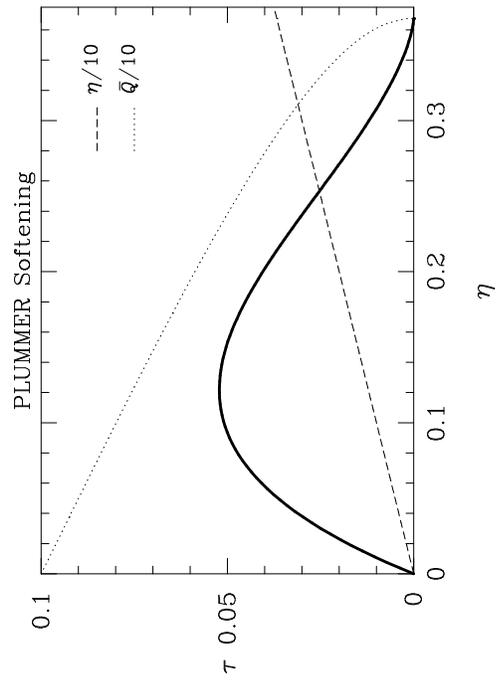
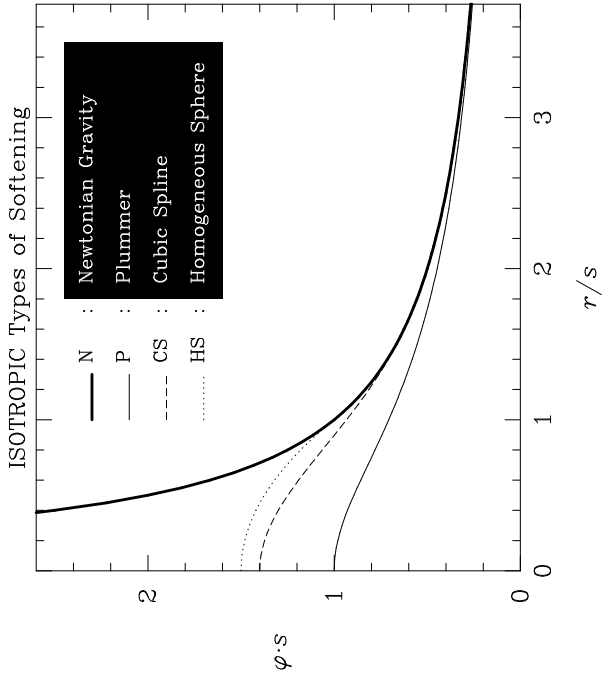
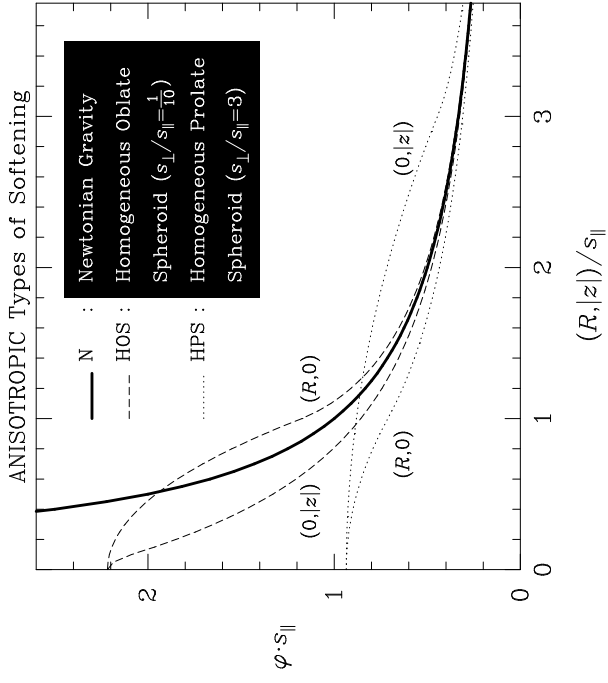
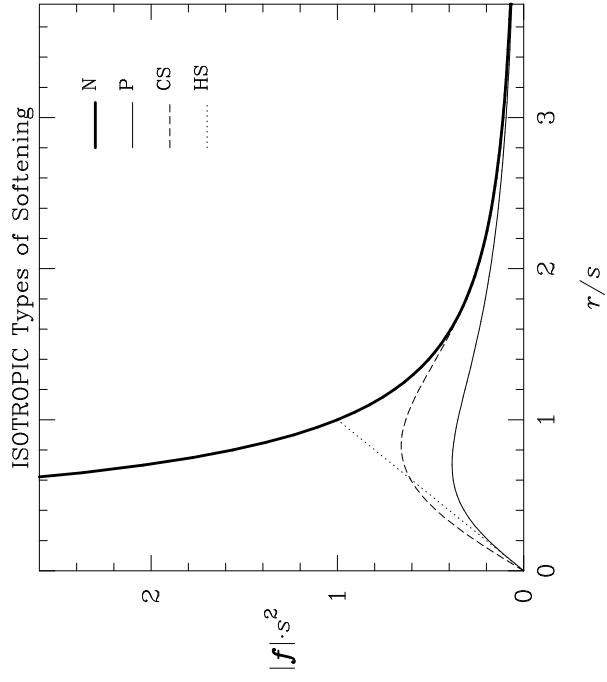
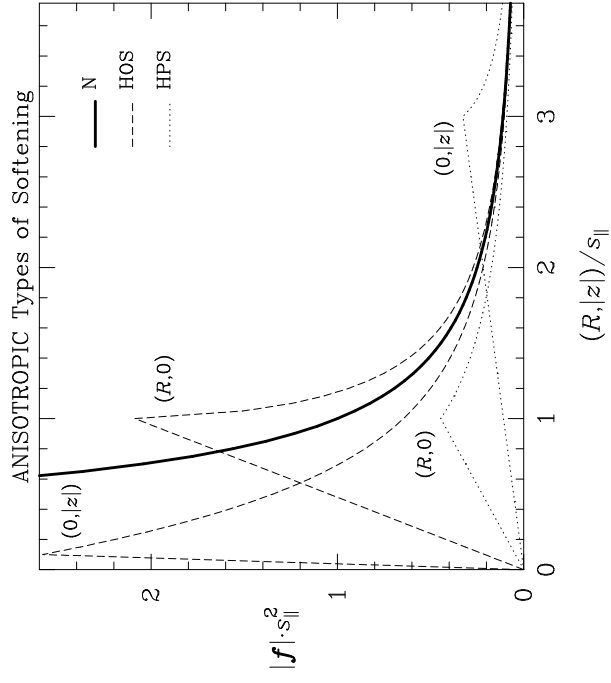
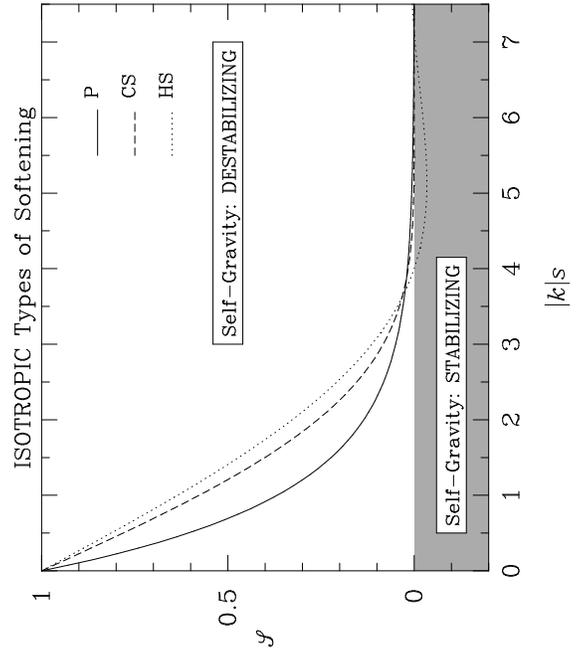
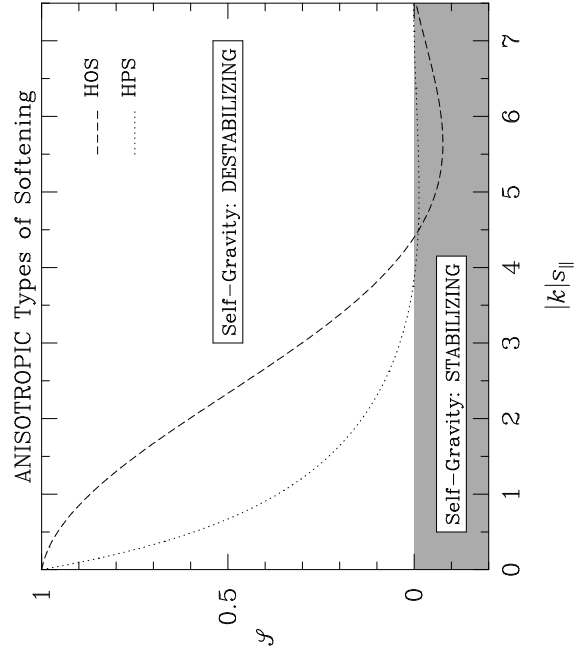
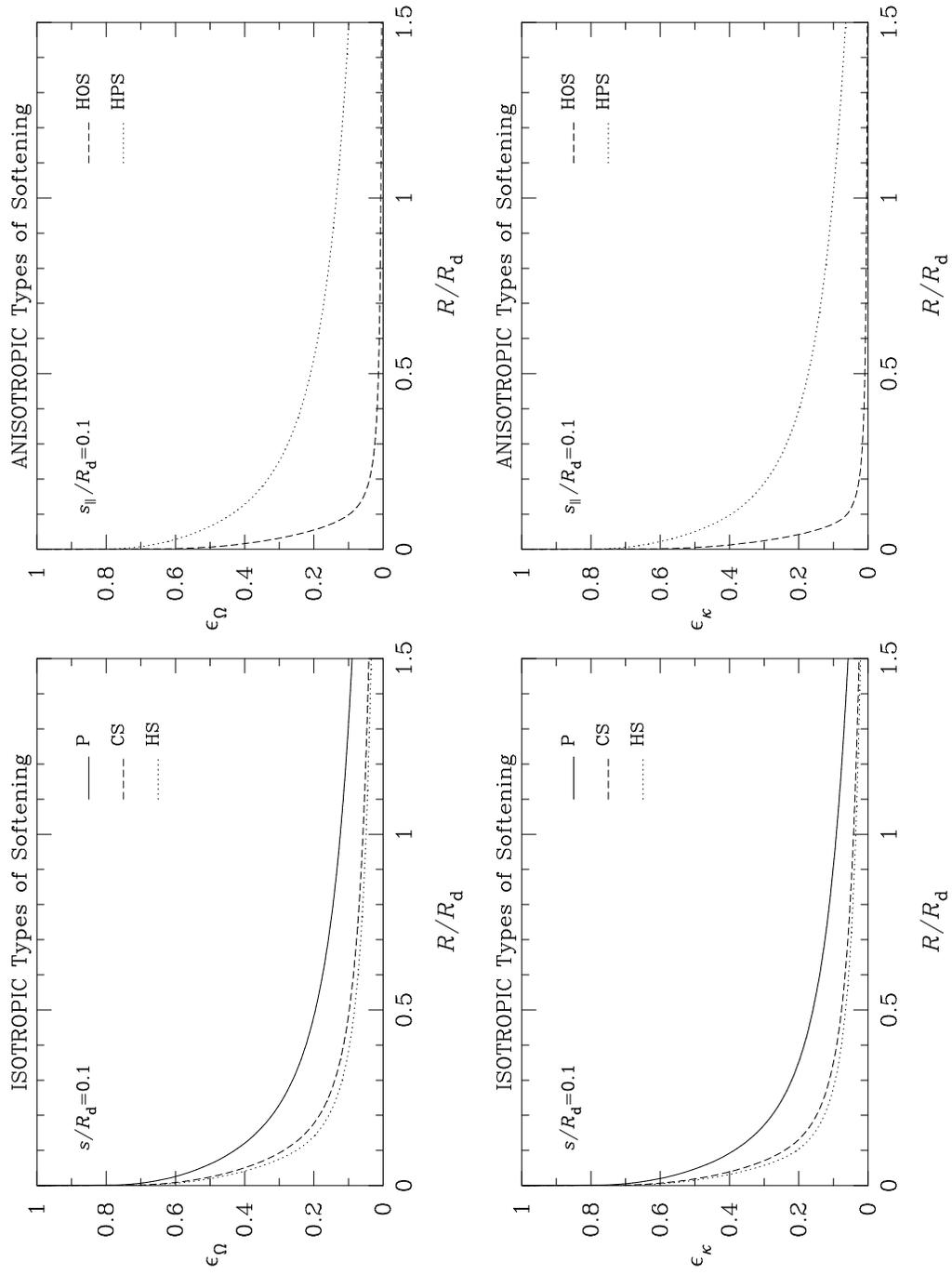
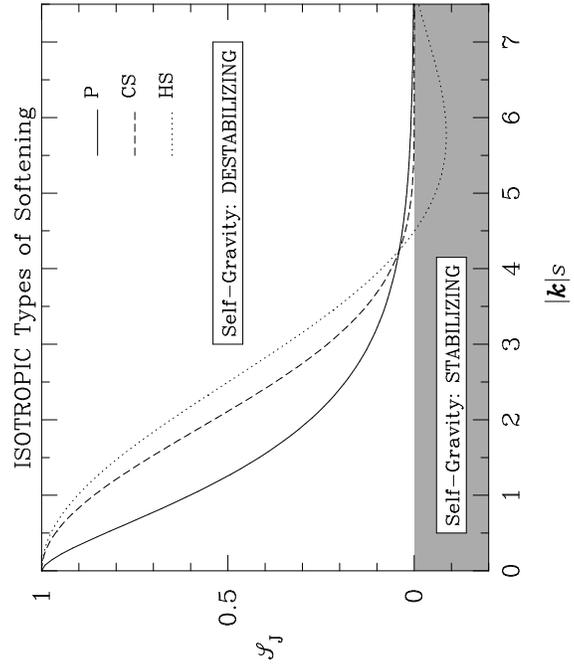
\end{document}